\documentclass[utf8]{frontiersSCNS} % for Science, Engineering and Humanities and Social Sciences articles
\usepackage{url,hyperref,lineno,microtype,subcaption}
\usepackage[onehalfspacing]{setspace}
\usepackage{hyperref}
\hypersetup{
    colorlinks=true,
    linkcolor=red,
    citecolor = blue,
    filecolor=cyan,      
    urlcolor=magenta,
}
\chardef\us=`\_

\def\keyFont{\fontsize{8}{11}\helveticabold }
\def\firstAuthorLast{Pant {et~al.}} %use et al only if is more than 1 author
\def\Authors{V. Pant\,$^{1,2,3,*}$, S. Majumdar\,$^{4}$, R. Patel\,$^{1,4}$, A. Chauhan\,$^{5}$ D. Banerjee\,$^{1,4,6}$, N. Gopalswamy\,$^{7}$}
\def\Address{$^{1}$ Aryabhatta Research Institute of Observational Sciences, Nainital, Uttarakhand 263002\\
$^{2}$ Instituto de Astrofis\'ica de Canarias, E-38205 La Laguna, Tenerife, Spain.\\
$^{3}$ Departamento de Astrofis\'ica, Universidad de La Laguna, E-38206 La Laguna, Tenerife, Spain. \\
$^{4}$ Indian Institute of Astrophysics, Bangalore - 560034, India.\\
$^{5}$ Harish-Chandra Research Institute, Allahabad - 211 019, India.\\
$^{6}$ Center of Excellence in Space Sciences, IISER Kolkata, India.\\
$^{7}$ NASA Goddard Space Flight Center, Greenbelt, MD, USA.}
\def\corrAuthor{V.~Pant}

\def\corrEmail{vaibhav@aries.res.in}

\begin{document}
\onecolumn
\firstpage{1}

\title[Power laws in CME widths]{Investigating width distribution of slow and fast CMEs in solar cycles 23 and 24} 

\author[\firstAuthorLast ]{\Authors} %This field will be automatically populated
\address{} %This field will be automatically populated
\correspondance{} %This field will be automatically populated

\extraAuth{}% If there are more than 1 corresponding author, comment this line and uncomment the next one.
%\extraAuth{corresponding Author2 \\ Laboratory X2, Institute X2, Department X2, Organization X2, Street X2, City X2 , State XX2 (only USA, Canada and Australia), Zip Code2, X2 Country X2, email2@uni2.edu}

\maketitle

\begin{abstract}

Coronal Mass Ejections (CMEs) are highly dynamic events originating in the solar atmosphere, that show a wide range of kinematic properties and are the major drivers of the space weather. The angular width of the CMEs is a crucial parameter in the study of their kinematics. The fact that whether slow and fast CMEs (as based on their relative speed to the average solar wind speed) are associated with different processes at the location of their ejection is still debatable.  Thus, in this study, we investigate their angular width to understand the differences between the slow and fast CMEs. We study the width distribution of slow and fast CMEs and find that they follow different power law distributions, with a power law indices ($\alpha$) of -1.1 and -3.7 for fast and slow CMEs respectively. To reduce the projection effects, we further restrict our analysis to only limb events as derived from manual catalog and we find similar results. We then associate the slow and fast CMEs to their source regions, and classified the sources as Active Regions (ARs) and Prominence Eruptions (PEs). We find that slow and fast CMEs coming from ARs and PEs, also follow different power laws in their width distributions. This clearly hints towards a possibility that different mechanisms might be involved in the width expansion of slow and fast CMEs coming from different sources.These results are also crucial from the space weather perspective since the width of the CME is an important factor in that aspect.

\tiny
 \keyFont{ \section{Keywords:} Coronal Mass Ejections, Corona, Kinematics, Space Weather} %All article types: you may provide up to 8 keywords; at least 5 are mandatory.
\end{abstract}

\section{Introduction}
\label{sec1}

Coronal mass ejections (CMEs) consist of plasma and magnetic field that are expelled from the solar atmosphere into the heliosphere at speeds which can range from 100 -- 3000 km $s^{-1}$ \citep{2004JGRA..109.7105Y,2004ASSL..317..201G,2010nspm.conf..108G,2011JASTP..73..671M,2012LRSP....9....3W}. They appear as bright, white-light features moving outward in the coronagraph field of view (FOV) \citep{1984JGR....89.2639H,1996Ap&SS.243..187S}. Though, early observations of CMEs date back to 1970s \citep{1971PASAu...2...57H}, it is \citet{1973spre.conf..713T} who first observed CMEs in the coronagraph images \citep[see recent review by][on the history of CMEs]{2016GSL.....3....8G}. Since the launch of the Large Angle and Spectrometric Coronagraph (LASCO) \citep{1995SoPh..162..357B} on the {\it Solar and Heliospheric Observatory (SOHO)} and Sun Earth Connection Coronal and Heliospheric Investigation (SECCHI) \citep{2008SSRv..136...67H} on the {\it Solar Terrestrial Relation Observatory (STEREO)}, CMEs are being routinely monitored.

CMEs are the major drivers of space weather, as they are capable of producing shock waves and interplanetary disturbances \citep{1991JGR....96.7831G,1993JGR....9818937G}, where the height of formation of the shock wave can often be estimated from the radio observations \citep{gopal2013}. Recently, \cite{vourlidas2020} have outlined the role of radio observations of CMEs during different stages of the CME eruption and its subsequent propagation in the heliosphere. Thus it is important to understand the kinematics of CMEs. CMEs during their radial propagation,  have been known to follow a three phase kinematic profile \citep{Zhang2001,Zhang_2006,2012LRSP....9....3W}. During their propagation, they interact with the ambient solar wind and experience a drag, which leads to a decreasing or constant speed in the later stages of their propagation \citep{2012LRSP....9....3W}. This average solar wind speed reportedly divides the CMEs into slow and fast \citep{nat2000}.  CMEs are also known to be associated with active regions and eruptive prominences \citep{2012LRSP....9....3W,2001ApJ...561..372S}. These two classes of source regions of CMEs tend to associate CMEs to two distinct classes \citep{1983SoPh...89...89M}. Later, \citet{Sheeley1999JGR} used the data from LASCO and confirmed this classification by suggesting that there are two dynamical classes of CMEs, which are gradual and impulsive CMEs. The former are slower and are preferentially associated with eruptive prominences, whereas, the latter CMEs are faster and are mostly associated with flares and active regions. So, it seems that these two classes of source regions of CMEs also tend to segregate CMEs into being gradual or impulsive events. The most intriguing question in this context is whether there are two physically different processes that are involved in the launch of these slow and fast CMEs or whether they belong to a dynamical continuum with a single unified process, the answer to which is still not clear \citep[also see][]{2012LRSP....9....3W}.

Apart from their radial propagation, CMEs are also known to exhibit lateral expansion that leads to an increase in their angular width as they propagate outwards \citep{Kay_2015,2020A&A...635A.100C,2020ApJ...899....6M} and that it is the Lorentz force at their source region that is closely responsible for translating and expanding them \citep{Subramanian_2014}. In this regard, \citet{2017ApJ...849...79Z} reported on the importance of the angular width of a CME, in determining whether the corresponding interplanetary CME and the preceding shock will reach Earth. \citet{2017ApJ...848...75L} reported on the importance of studying the expansion in slow CMEs on their ability to drive shocks. The width of the CME also sheds light on the source region of the CME it is coming from.  \citet{2007ApJ...668.1221M} showed that the strength of the magnetic field of the source region flare arcade producing a CME can be estimated from the final angular width of the CME and the angular width of the flare arcade. Recently \citet{2020ApJ...899....6M} connected 3D profiles of width evolution to the 3D acceleration profiles of slow and fast CMEs and found that the vanishing of the initial impulsive acceleration phase and the ceasing of width expansion phase both tend to occur in a height range of 2.5-3 R$_\odot$, thus showing the observational evidence of the height of impact of Lorentz force on CME kinematics. So, it is evident that the width of a CME is an essential ingredient in the understanding of their kinematics, and is also an important parameter for the consideration of their space weather impact. Furthermore, since the width largely influences the kinematics of CMEs, it is still not known whether we observe any differences in the angular width distribution of slow and fast CMEs originating from different source regions. 

It has also been reported that the width distribution of CMEs follow a power law \citep{2006ApJ...650L.143Y,2009ApJ...691.1222R,2014ApJ...795...49D}. A study of the statistical distribution of a physical parameter sheds light on the underlying physics of it, and a presence of power law in the distribution of a quantity indicates the presence of Self-Organized Criticality \citep{PhysRevLett.59.381}. The presence of power laws, and hence Self-Organized Criticality (SOC) in nature have become evident in the last few years in many different areas and astrophysical phenomena \citep[and references therein]{2018SSRv..214...55A}. The presence of power laws in solar astrophysics, in the global energetics of solar flares, has also been reported by \citet{2016ApJ...831..105A}. Thus, a study of the distribution of the angular width of CMEs should provide important clues in understanding the physical mechanisms responsible for expanding the CMEs. Recently, \citet{bidhu_ss} studied the distribution of width of CMEs during the maximum phase of solar cycle 23 and 24. \citet{2014ChA&A..38...85M} studied the distribution of CME width and its comparison with the phase of sunspot number in solar cycle 23. Inspite of these studies, whether the slow and fast CMEs follow different width distributions or whether there is any imprint of the source regions on the width distribution of these slow and fast CMEs, is still not properly understood. Thus it is worth looking at the width distribution of slow and fast CMEs and also if there is any imprint of the source region of these two dynamical classes on their width distribution. 

Motivated by the above findings and the deficit in our understanding of these slow and fast CMEs,  we in this work study the width distribution of slow and fast CMEs that occurred during different phases of cycle 23 and 24. We outline the data sources and the working method in Section \ref{sect2}, followed by our results in Section \ref{sect3}. Finally, we outline our main conclusions from this work in Section \ref{discussion}.

%\subsection{Power laws in solar physics}

\section{Data and Method}
\label{sect2}
\subsection{Data Source}

We use the data from CDAW catalogue for the analysis presented in this paper. The CDAW catalog lists the properties of CMEs detected manually \citep{2004JGRA..109.7105Y,2009EM&P..104..295G} in SOHO/LASCO images. The work on source region identification and segregation is done with the images taken by {\textit{Atmospheric Imaging Assembly}} (AIA) on-board {  \textit{Solar Dynamics Observatory}} (SDO) \citep{aia} and the Extreme Ultraviolet Imager (EUVI) on-board STEREO (for details, refer to Section \ref{source_identification}) .
 
\subsection{Event Selection}
 
We have selected the CMEs from the CDAW catalogue that have occurred during different phases of solar cycle 23 and 24.  For the analysis presented in this work, we first remove the ``very poor" CMEs from the CDAW catalog. \citet{2014ApJ...784L..27W} reported that the detection of ``very poor" CMEs are based on the discretion of manual operators, we discard such CMEs in order to remove any bias from our analysis. It should be noted that some of the ``very poor" CMEs may be the real CMEs but we remove them from the analysis because there are large errors in the measurement of the properties of such CMEs. Furthermore, we impose a lower threshold of 30$^{\circ}$ on CME width to remove narrow CMEs. \citet{2008AnGeo..26.3103Y} and \citet{2010ASSP...19..289G} have reported that there exists a discrepancy in the detection of the number of CMEs with width $<$ 30$^{\circ}$ when both CACTus and CDAW catalogs were compared. Also \cite{narrow_cmes} studied the statistical properties of narrow CMEs, and reported that they do not form a subset of normal CMEs and have different acceleration mechanism. In addition to a lower threshold, we also apply an upper threshold of 180$^\circ$ on the width because such CMEs mostly suffer from projection effects and thus the width estimation will be affected. It is worth noting that CMEs with width between 30$^\circ$ and 180$^\circ$ also suffer from projection effects. In order to reduce projection effects, we also use limb CMEs (whose source regions were found within 30$^{\circ}$ of the limb) for the analysis. The selection criteria for the limb CMEs is reported in \citet{2014GeoRL..41.2673G}.
 \vspace{1em}
\subsection{Segregation of CMEs into slow and fast }

After shortlisting the CMEs based on the above selection criteria, we segregate the CMEs as slow and fast based on their speeds. CMEs are usually classified as slow and fast relative to the speed of the solar wind. The slow solar wind typically has speeds less than 400 km s$^{-1}$ while the fast solar wind has speeds greater 400 km $s^{-1}$ \citep[see,][]{2006LRSP....3....2S}. Therefore, 400 km s$^{-1}$ can be taken as the average solar wind speed for a long term statistical study. We classify CMEs with speeds less than 300~km~s$^{-1}$ as slow CMEs and those with speeds greater than 500~km s$^{-1}$ as fast CMEs. We consider CMEs with speeds between 300~km s$^{-1}$ to 500~km s$^{-1}$ as intermediate CMEs, as they cannot be strictly categorised as either slow or fast CMEs because of uncertainties in speed measurements. It is worth noting at this point that the speeds of CMEs listed in the CDAW catalog are the speeds with which the leading edge of a CME propagates. Table~\ref{table1} lists the number of fast and slow CMEs in the CDAW catalog.
\vspace{1em}
\subsection{Segregation of the slow and fast CMEs on the basis of their source region}
\label{source_identification} 
After segregating the CMEs as slow and fast, we search their source regions on the solar disk. For this we included a total of 1064 events that occurred during 2000 to 2002, 2008, 2009, and also from 2012 to 2014, which covers the maxima of cycle 23, the minima of cycle 24 and the maxima of cycle 24. Here, we have followed a similar method as reported in \citet{2020ApJ...899....6M} and using the JHelioviewer software \citep{jhelio2,jhelio} to back-project the CMEs onto the solar disk. We further segregated the identified source regions of the CMEs into two broad categories, namely (i) Active Regions(ARs) and (ii) Prominence Eruptions(PEs). We define them as follows:

\begin{figure}[ht]
\centering
 \includegraphics[trim=300 20 300 20, clip, scale=0.2]{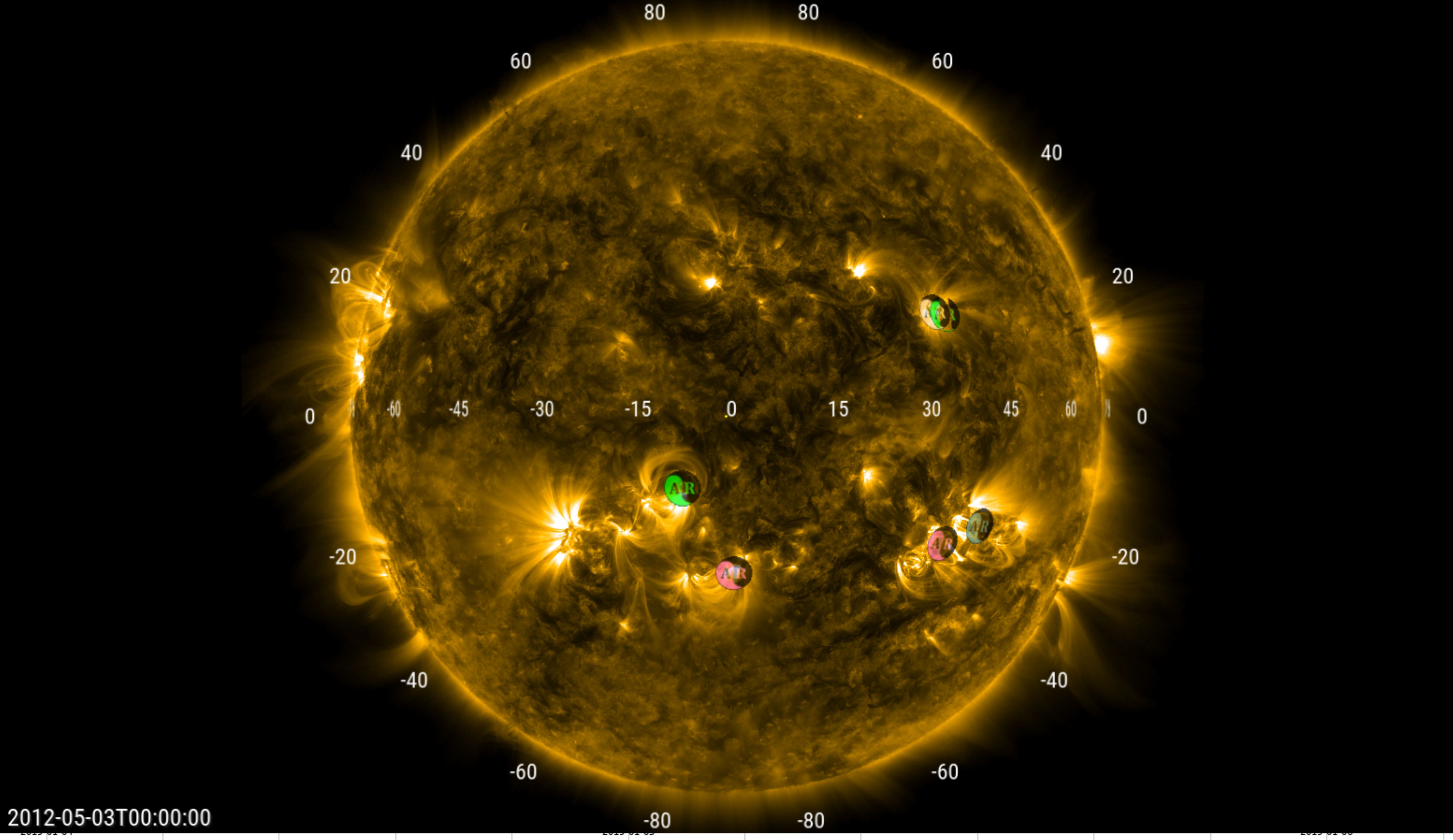}
 \includegraphics[trim=300 20 300 20, clip, scale=0.2]{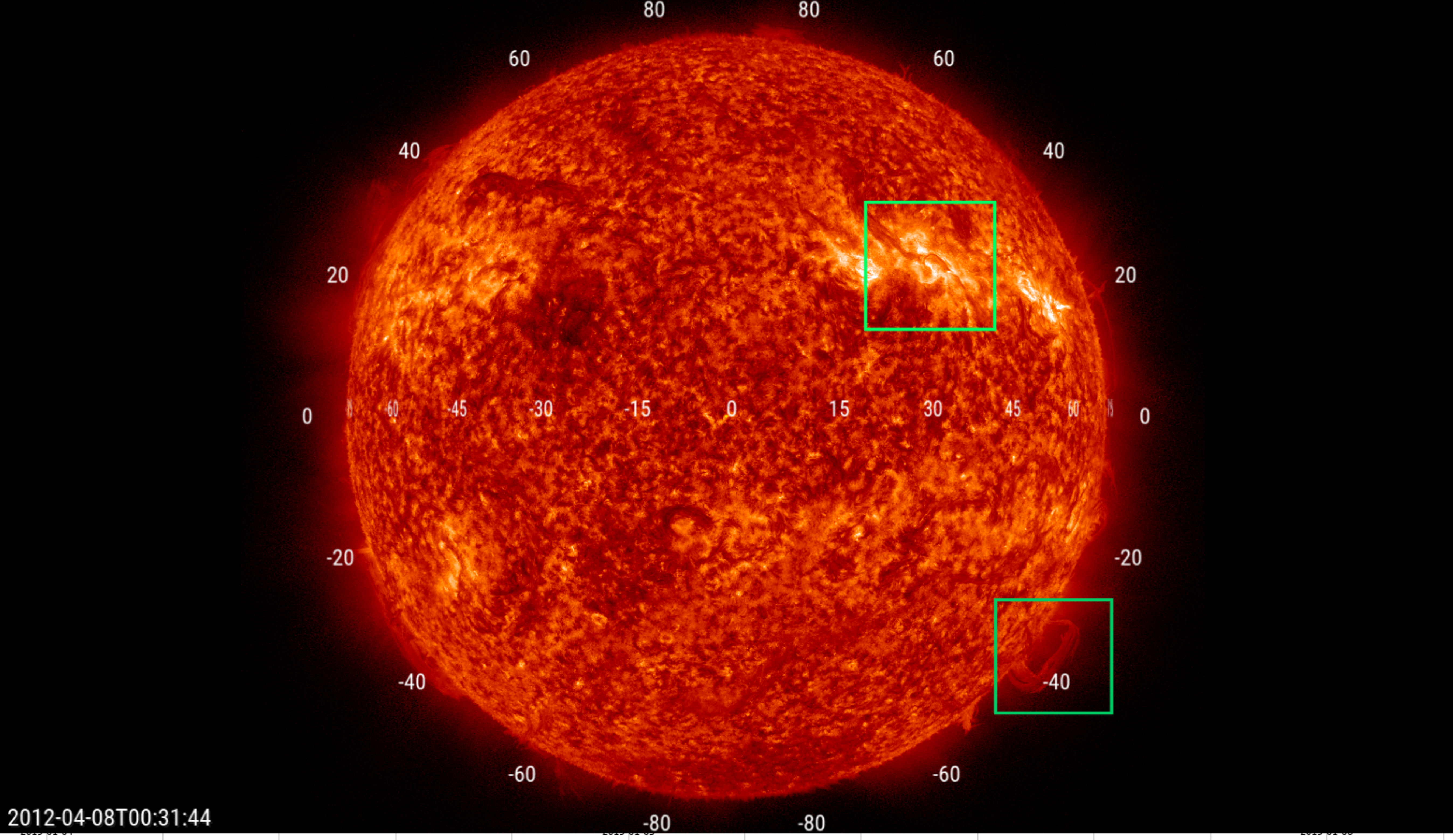}
  \caption{(Left)Active regions (ARs) as observed in the corona on 03-05-2012. The image is taken by AIA 171 \AA. The ARs (pointed out in the figure by circular marks with AR written on them), are identified as bright regions, being hotter and denser than the background coronal plasma, showing strong magnetic activity. (Right) Prominence Eruptions (PEs, enclosed in green box) as observed in the corona on 08-04-2012. This image is taken by AIA 304 \AA. The dark strands of plasma as seen projected against the disk of the sun are filaments (at a position angle of 40$^{\circ}$ and 300$^{\circ}$ in this image), whereas the same object is termed as a prominence when observed at the solar limb (at a position angle of 235$^{\circ}$ in this image).}
   \label{sr}
 \end{figure}
Active Regions(ARs) (refer Figure~\ref{sr}) are prime features making up only a small fraction of the total surface area of the Sun, but harbouring most of the solar activity  \citep{2012LRSP....9....3W}. These being areas of strong magnetic field, are predominantly hotter and denser than the background coronal plasma, producing bright emission in the soft X-ray and extreme ultraviolet regions. In order to identify ARs, we look into the images taken by (i) the Extreme Ultraviolet Imaging Telescope \citep[EIT;][]{EIT} (195 \AA) onboard SOHO, (ii) Atmospheric Imaging Assembly \citep[AIA;][]{aia} (171 \AA, 193 \AA) onboard SDO, and also (iii) the Extreme Ultraviolet Imager \citep[EUVI;][]{euvi}(171 \AA, 195 \AA) onboard STEREO SECCHI. We also detect the ARs by the Active Region numbers provided by the National Oceanic and Atmospheric Administration (NOAA) as supplied by Space Weather Event Knowledgebase (SWEK) using Jhelioviewer\footnote{https://www.jhelioviewer.org/} \citep{jhelio} . 
%\textbf{\textcolor{red}{ We also look for possible signatures of re-alignment of the magnetic field topology that leads to outward moving structures in the lower corona while seen in running difference images.}}

Prominences (refer Figure~\ref{sr}) are cool dense material (8000 K) embedded in the hotter corona, observed as an emission feature when seen at the solar limb, and an absorption feature when seen projected against the background hotter corona (termed as filament) \citep{Gilbert_2000}. We classify an eruption as a prominence eruption if we see a strong radial component of motion away from the solar surface where all or some of the prominence material is seen to escape the gravitational field of the Sun. For a filament eruption (we include these in the same category with prominence eruption) we either looked for tangential motion across the solar surface with a subsequent eruption, or simply by observing any disappearance of the filament in the subsequent images with a transient coronal manifestation following it (also refer \citet{1987SoPh..108..383W}). It is also important to note that, there can be possibilities of failed eruptions \citep{Njoshicmefailederupt2013ApJ}, CME-jet interactions \citep{2019ApJ...881..132D,2020SoPh..295...27S}, and CME-CME interactions \citep{Njoshi2013,cme-cme_interations} which can influence the kinematic properties. Furthermore, CMEs are also observed to be associated with coronal jets, minifilaments, etc \citep{2015ApJ...813..115L,2019SoPh..294...68S,2019ApJ...881..132D}. It is also not properly understood whether such CMEs can be categorised as the active region eruptions and prominence eruptions or classified as a different category. Thus, in the present work, we exclude such events. Furthermore, CMEs are also related to the active region filaments, however, due to the lack of comparable statistics with the other two classes (ARs and PEs), we have excluded such events from our analysis. To detect PEs, we look into the images taken by, (i) EIT 195 \AA, 304 \AA, (ii) AIA 304 \AA, and (iii) EUVI 304 \AA. Finally, it should be borne in mind that kinematic properties of CMEs largely depend on the overlying field strength and decay index \citep{Njoshicmefailederupt2013ApJ,2012ApJ...761...52X}. In this work we do not consider these effects into account.
%It is worthwhile to note that we do-not include Active Prominences (which we define as PEs with one or more foot-points being connected to ARs, (see \citet{2001ApJ...561..372S}) in this category. We do-not include them in this work due to poor statistics.

For a spatial association \citep[see][]{Gilbert_2000,2020ApJ...899....6M} between a source region and a subsequent CME, we require that the latitude of the source region to be around $\pm$ 30$^{\circ}$ to that of the PA of the center of the CME as reported in the CDAW catalogue. In the case of a filament eruption, due to larger uncertainty of its spatial location, we look for an erupting filament around $\pm$ 40$^{\circ}$ around the PA of the CME converted to equivalent apparent latitude ($lat_{PA}$) by the following relation:

\begin{equation}
    lat_{PA} = 90-PA \quad [0 \leq PA \leq 180] \qquad
    lat_{PA} = PA-270 \quad [180 < PA \leq 360].
\end{equation}

For a temporal association we consider source region to erupt or show radially outward movement in the above latitude window in a time interval of at least 30 minutes before the first appearance of the leading front in the LASCO C2 field of view.

 \section{Data analysis and results}
\label{sect3}

 \subsection{Width distribution of CMEs}
\begin{figure}[ht]
\centering
 \includegraphics[scale=0.75]{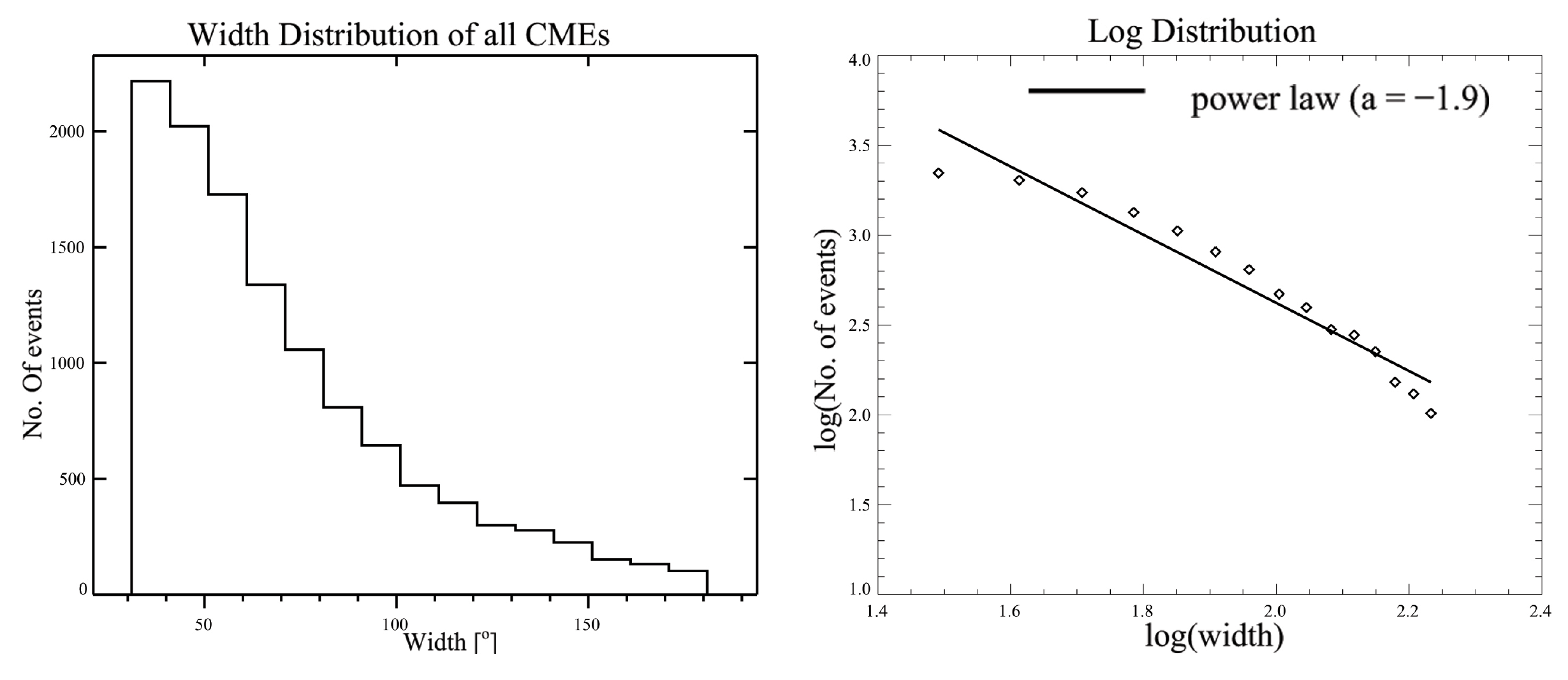}
  \caption{Width distribution of all CMEs (excluding "very poor" events) during solar cycle 23 and 24 using the CDAW catalogue. The black line corresponds to a power law fit to the width distribution where $\alpha$ is the power law index. }
   \label{width_all_events}
 \end{figure}

\begin{figure}[ht]
\centering
 \includegraphics[scale=0.75]{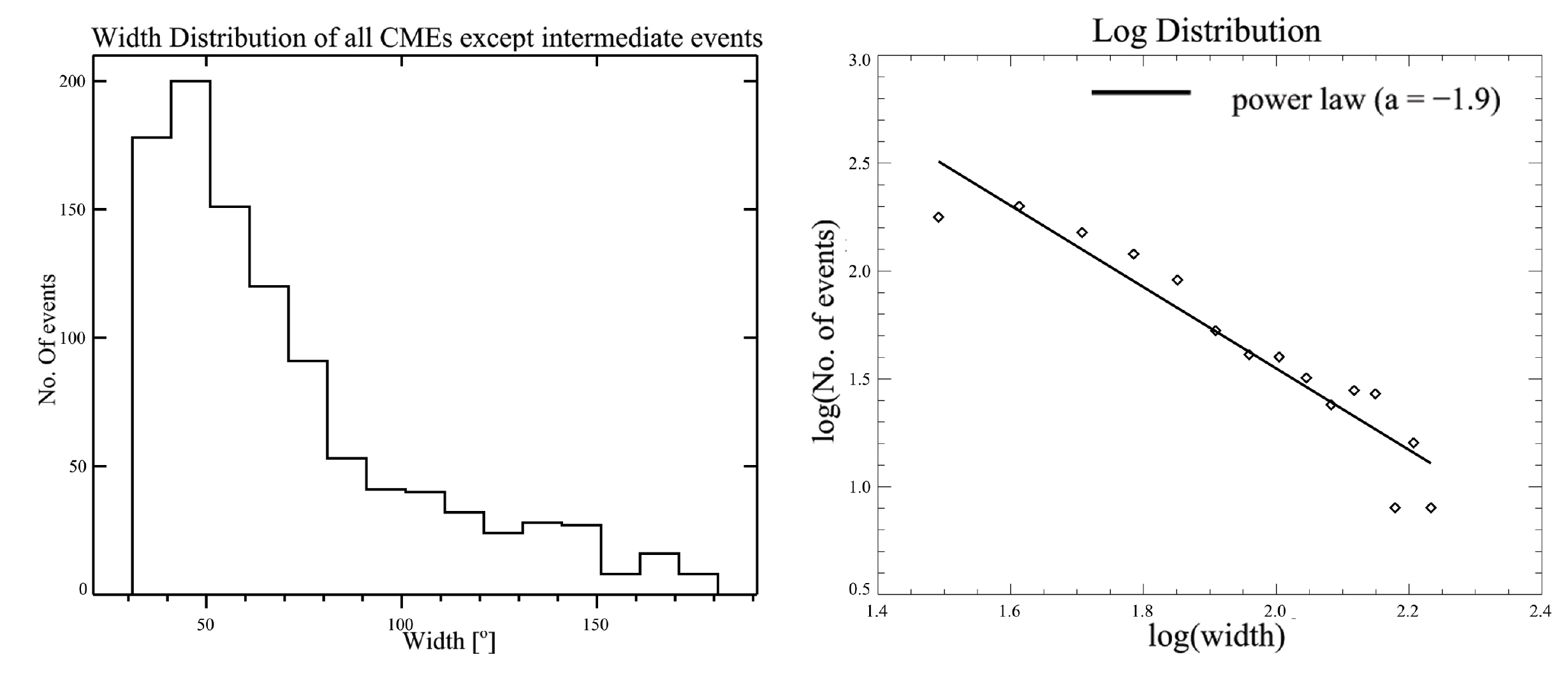}
  \caption{Width distribution of CMEs (excluding the intermediate and "very poor" events) during solar cycle 23 and 24 using the CDAW catalogue. The black line corresponds to a power law fit to the width distribution where $\alpha$ is the power law index. }
   \label{width_all}
 \end{figure}

Recently, it has been reported that the CMEs evolve non self-similarly in the inner corona \citep[see][]{2020ApJ...899....6M,2020A&A...635A.100C}. In the CDAW catalog, the width of a CME is defined as the maximum angle subtended by a CME on the center of the Sun when the CME enters the LASCO C3 field of view (FOV)  where the width appears to approach a constant value \citep{2004ASSL..317..201G}. To investigate the width distribution of CMEs during solar cycle 23 and 24, we fit a power law to it as follows:

\begin{linenomath*}
\begin{equation}
\label{eq1}
N (W)=CW^{\alpha},
\end{equation}
\end{linenomath*}
where N is the number of CMEs with width $W$,  $\alpha$ is the power-law exponent, and $C$ a constant.
In Figure~\ref{width_all_events} we plot the histogram (left panel) of the width distribution and the width distribution in log scale (right panel) with the power law fit for all CMEs excluding the "very poor" events as mentioned in the CDAW catalogue. It is worth noting that after removing ``very poor" CMEs from our analysis, we believe that we have reduced the bias introduced by manual operators, respectively. We find a power law index of -1.9. In order to understand the goodness of fit we perform the Kolmogorov Smirnov (KS) test where the KS distance \citep[see][]{2007arXiv0706.1062C} which is defined as the maximum distance between the empirical distribution function of the sample and cumulative distribution function of the assumed expression, is minimum for the distribution which fits the data best with a corresponding high p-value that gives the probability confidence. Here we find the KS distance and p-value as 0.13 and 0.99. It is worth noting from Figure~\ref{width_all_events}, that the width distribution is not fitted well by a single power law. This serves as a motivation for us to investigate power laws segregating fast and slow CMEs\\
Since we aim to understand the width distribution of slow and fast CMEs, we next remove the intermediate events from our study sample, as such events are neither fast nor slow (refer section 3.2). We again study the width distribution of all events except the "very poor" and the intermediate events (Figure~\ref{width_all}). We find that after removing the intermediate events, we still get the same power law index of -1.9 with the KS distance and p-value to be the almost same. Thus we ensure that there is no bias introduced in the estimation of the power law index of widths distribution by rejecting the intermediate events from our sample. The graphical fitting (GF) of the data points used above is not the best method to estimate the power-law, especially when number of data points is small  \citep{2016SoPh..291.1561D}. Therefore, we also use maximum likelihood estimate (MLE) fitting method to derive the power-law index, $\alpha$. Using MLE, we get the power law index to be -1.6 which again remains the same for CMEs with or without the intermediate events. 

Now that we have ensured that the exclusion of the intermediate events does not affect our study, we try to understand now if the slow and fast CMEs follow different power laws in their width distribution. 

\vspace{1em}
\subsection{Width distribution of slow and fast CMEs}

\begin{figure}[ht!]
\centering
\includegraphics[scale=0.40,angle=0]{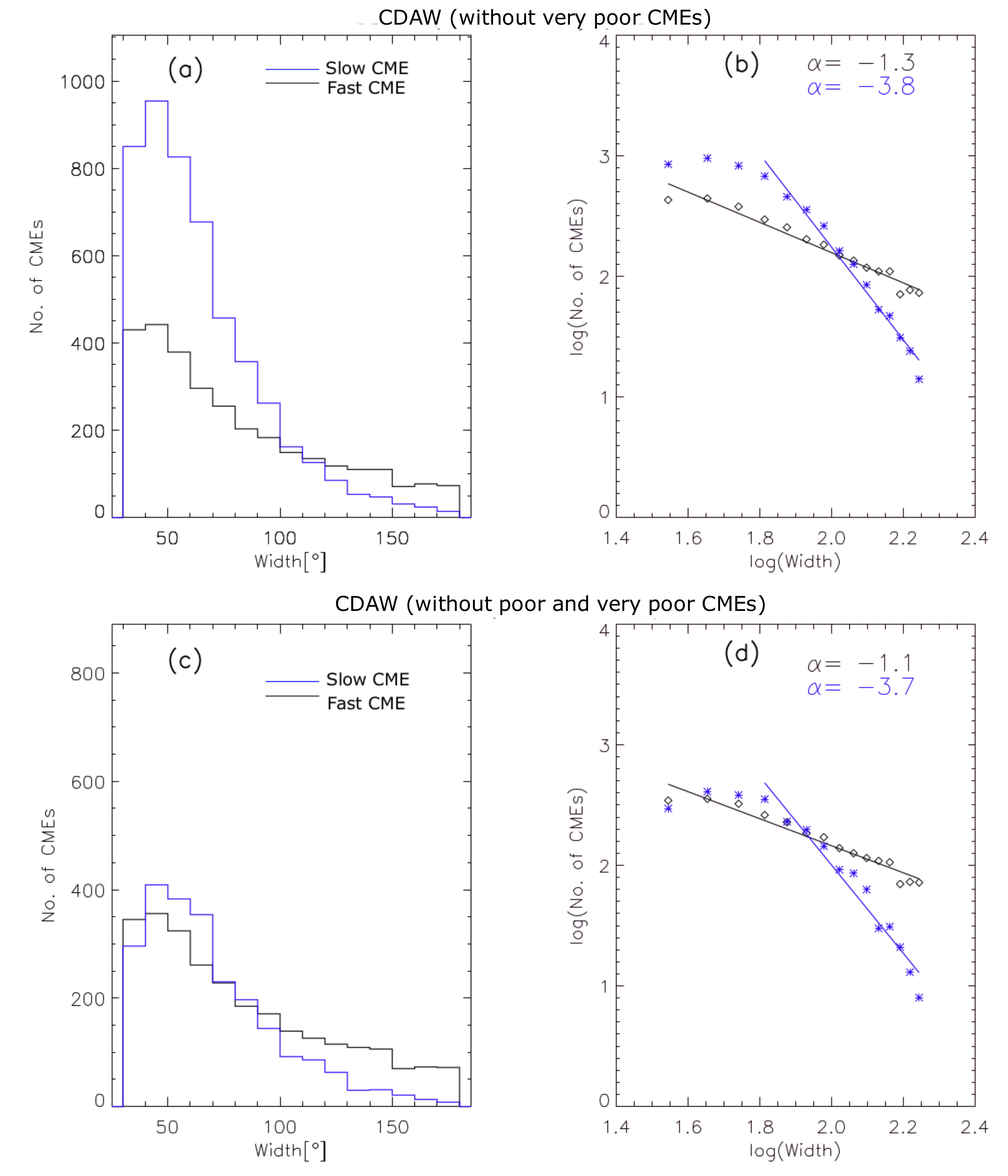}
\caption{(a): Width distribution and, (b): log distribution of width for slow and fast CMEs using CDAW catalog after excluding ``very poor'' CMEs. The best fit straight line to the data points of fast CMEs is overplotted in black.  The best fit straight line to the data points (except first three points) of slow CMEs is overplotted in blue. (c) and (d) are the same as (a) and (b) but exclude ``poor'' and ``very poor'' CMEs from the CDAW catalog. }
\label{c7f1}
\end{figure}

In this section we study the width distribution of slow and fast CMEs separately. We use the same power law distribution as mentioned in equation \ref{eq1}. Figure~\ref{c7f1} (a) shows the histograms of the width distribution of fast (black line) and slow (blue line) CMEs using CDAW catalog after excluding the ``very poor" CMEs and the intermediate CMEs. Figure~\ref{c7f1} (b), shows the width distributions in log scale. Black and blue lines represent the best fit power-laws obtained for the fast and slow CMEs, respectively using least square fitting method. We estimate $\alpha$ as -1.3 and -3.8 for fast and slow CMEs respectively. Thus we find that the slow and fast CMEs follow different power laws in their width distribution, but to get a better confidence on our result, we again perform power law fitting by MLE.

It should be noted that the number of slow CMEs in CDAW catalog with widths less than 70$^{\circ}$ flattens. This could partially be due to the observational limitations of C2 coronagraph or human subjectivity or both. The minimum width for the graphical fitting is decided by visually inspecting the distribution and choosing the width beyond which the distribution is supposedly following a power-law.
Thus, we fit the tail of the width distribution of slow CMEs with a power-law. Later, we used MLE to derive the minimum width beyond which the distribution is best represented by the power law. 

%To fit a straight line to the width distribution of slow CMEs, we use the width greater than 70$^\circ$ ({\it i.e,} the tail of the width distribution).
We performed MLE fitting in two different ways. First we set the minimum width value, $W_{m}$ as 30$^\circ$ for both fast CMEs and slow CMEs and estimate the power-law index. Second, we derive the minimum width value,  $W_{d}$ for both fast and slow CMEs by minimising the KS distance and estimate the power-law index. The second method tells us that beyond  $W_{d}$, the data points best follow the power-law. Table~\ref{table1} lists the power indices estimated using two methods described above for fast and slow CMEs. We see that indeed the slow and fast CMEs follow distinctly different power laws in their width distribution. \\
%We find that the derived minimum width for slow CMEs is in agreement with the manually given minimum width for graphical fitting (GF). \\
\begin{table}[htb]
%\begin{center}
\footnotesize
\centering
\caption{Power--law indices of width distribution of fast and slow CMEs obtained using two different methods}  
\label{table1}
\begin{tabular}{cc||c|cc|cccc||c|cc|cccc}
%\begin{tabular}{cccccccccccccccc}
\hline
\multicolumn{1}{c}{Catalog} & \multicolumn{1}{c||}{Total CMEs} & \multicolumn{7}{c||}{Fast CMEs} & \multicolumn{7}{c}{Slow CMEs} \\
%\hline
    &   &  Total & \multicolumn{2}{c|}{GF} & \multicolumn{4}{c||}{MLE} & Total & \multicolumn{2}{c|}{GF} & \multicolumn{4}{c}{MLE}\\
 \hline
    &   &     & $W_{m}$ & $\alpha$ & $W_{m}$ & $\alpha_{m}$ & $W_{d}$ & $\alpha_{d}$ &   & $W_{m}$ & $\alpha$ & $W_{m}$ & $\alpha_{m}$ & $W_{d}$ & $\alpha_{d}$\\

%\hline	
%CDAW$^{\dagger}$ &19046&2554&40&-1.4&40&-1.3&66&-1.5&3967&80&-4.1&40&-2.28&89&-4.4\\
%CDAW$^{\ddagger}$&11329 & 2298  & 40&-1.2&40&-1.1&66&-1.4& 2016 &80&-3.9&40&-2&80&-3.8\\
%SEEDS&50955 &2023& 40&-2.8&40&-2.1&82&-3.6& 3019&80&-5.7&40&-3.2&87&-5.9\\
%Limb CMEs&531&233&40 &-0.2 &40 & -1& 68& -1&147&40 & -3.2& 40 & -2.51 &68 &-4.1\\
CDAW$^{\dagger}$ &19046&3031&30&-1.3&30&-1.13&66&-1.48&4925&70&-3.8&30&-1.8&89&-4.36\\
CDAW$^{\ddagger}$&11329 & 2680  & 30&-1.1&30&-1.01&66&-1.37& 2357 &70&-3.7&30&-1.53&80&-3.74\\
Limb CMEs&531&266&30 &-0.4 &30 & -1& 68& -1&169&70 & -3.8& 30 & -1.8 &68 &-4.16\\
%CACTus&18949&2045&30 &-1.6 &30 & -1.48& 70& -2&1956&70 & -4.8& 30 & -2.34 &88 &-5.71\\
\hline
\multicolumn{16}{l}{ $\dagger$ CMEs excluding ``very poor'' CMEs. $\ddagger$ CMEs excluding ``poor'' and ``very poor'' CMEs.} \\
\multicolumn{16}{l}{$W_{m}$ is the minimum width threshold used to fit datasets. $\alpha$ is the power index estimated using graphical fitting (GF) method.}\\
\multicolumn{16}{l}{$\alpha_{m}$ is the derived power index using the Maximum likelihood Estimate (MLE) by giving a minimum width threshold, specified by $W_{m}$.}\\
\multicolumn{16}{l}{$W_{d}$ is the derived minimum width by minimising the Kolmogorov-Smirnov (KS) distance.}\\
\multicolumn{16}{l}{$\alpha_{d}$ is the derived power index by applying MLE using minimum width threshold as $W_{d}$.}\\
\multicolumn{16}{l}{Limb CMEs are extracted from the CDAW catalog according to the criteria in \citet{2014GeoRL..41.2673G}. }

\end{tabular}
%\end{center}
\end{table}

\begin{figure}[ht!]
\centering
 \includegraphics[scale=0.45]{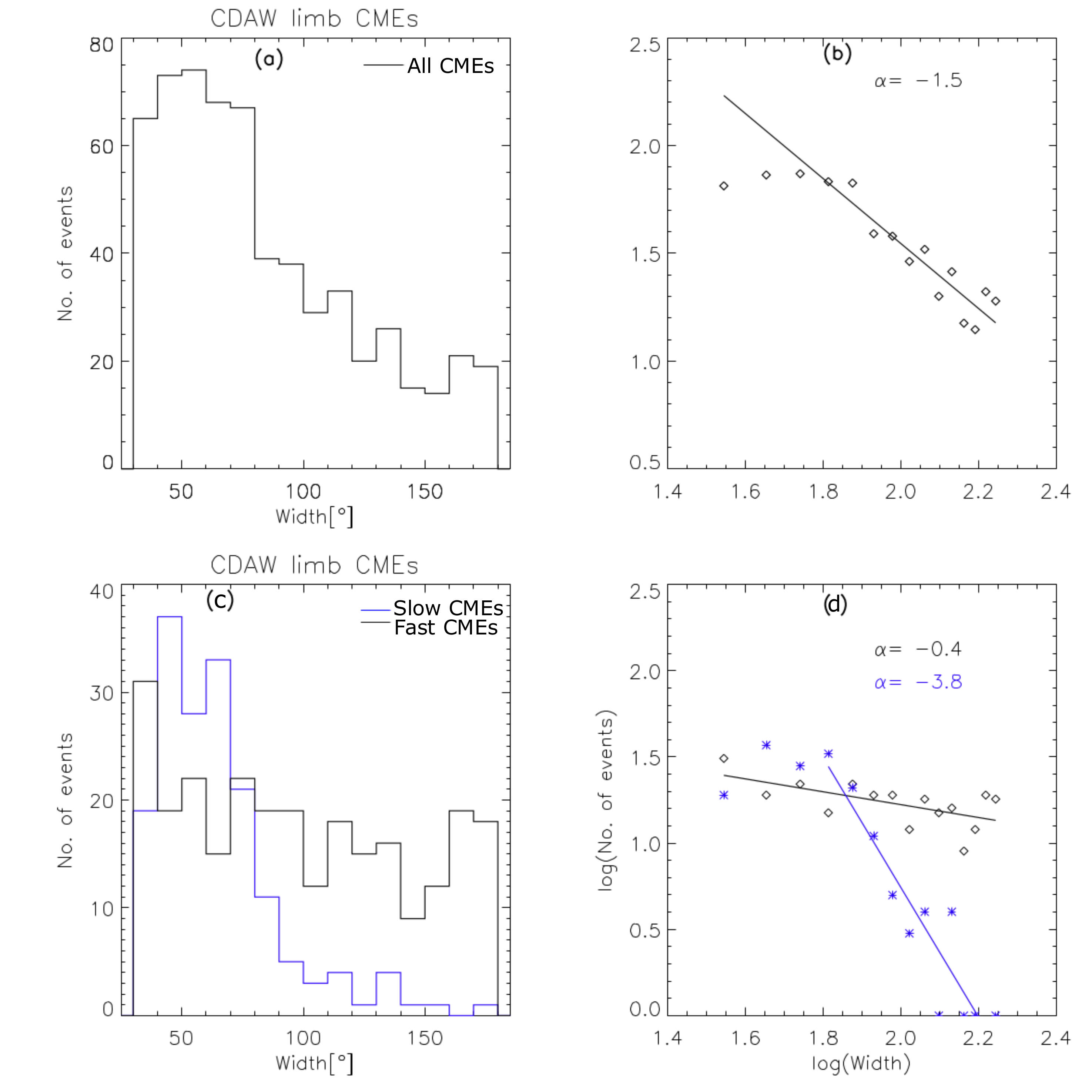}
  \caption{(a): Width distribution for all limb CMEs extracted from CDAW catalog without velocity thresholding. (b) is the log-distribution of limb CMEs. (c) and (d): Width distribution of slow and fast limb CMEs extracted from CDAW catalog.}
   \label{c7f3}
 \end{figure}

We also remove the CMEs labelled as ``poor" in the CDAW catalog to study their effects on the width distribution.  We find that removal of both ``poor" and ``very poor" CMEs has little effect on the width distribution and power-law indices (see, Figures~\ref{c7f1} (c) and (d) and Table~\ref{table1}) of slow and fast CMEs.
%We also find that while the power indices of fast and slow CMEs derived using CACTus and CDAW catalogs are in agreement with each other, those derived using SEEDS catalog deviate from both CDAW and CACTus using either of the two methods described above. We note that the fast CMEs beyond 140$\degree$ in SEEDS catalog are less than expected, which makes power index steeper for fast CMEs.\\
The speeds listed in the CDAW catalog are the projected speeds in the plane of sky measured at a fixed position angle. Therefore, applying a speed threshold uniformly to all CMEs introduces projection effects. In order to reduce projection effects, we also estimate the power-law for fast and slow limb CMEs in solar cycles 23 and 24. First we plot CMEs with all velocities whose widths fall between 30--180$^\circ$ (top panel of Figure~\ref{c7f3}). We estimate the power index of -1.5 and -1.19 using the GF and MLE methods for the minimum width threshold of  30$^\circ$. Then, we segregate fast and slow CMEs and estimate the power indices using the GF (bottom panel of Figure~\ref{c7f3}) and MLE methods (see Table~\ref{table1}). We find that the power indices for fast and slow limb CMEs are different from each other; they are different from the power index of non--limb fast and slow CMEs. One of the reasons for this discrepancy is the small number of slow and fast limb CMEs. From Table~\ref{table1}, we note that the fast and slow limb CMEs are 10--20 times less than fast and slow CMEs when non-limb events are also considered. To evaluate the goodness of fit we again estimate the KS distance. The KS distances for fast limb CMEs and all fast CMEs are 0.09 and 0.01, respectively.  The critical values of KS distances with 99\% of confidence limit for fast limb CMEs and all fast CMEs are 0.1 and 0.03, respectively. Smaller the KS distance, better is the fit. Similarly, KS distances for slow limb CMES and all slow CMEs are 0.04 and 0.02, respectively. Further, critical KS distance at 99\% of confidence limit for slow limb CMEs and all slow CMEs are 0.13 and 0.03, respectively.Thus the power indices corresponding to the limb CMEs differ from that of non-limb cases. We want to emphasize that the slow CME power-law is steeper than that of the fast CMEs in both cases, although the values may differ. However, the limb CME values may be closer to reality because of minimal projection effects. It should be noted that the results are consistent with the speed--width relation as reported in \citet{2014GeoRL..41.2673G}; where authors have reported wider CMEs tend to propagate faster than narrow CMEs. Also, we note that the KS distance of fast CMEs without``very poor" and intermediate events is of an order of magnitude better than KS distance estimated by fitting a single power law to all CMEs. This demonstrates that fast and slow CMEs are better represented by different power-laws than fast and slow CMEs combined. 
\vspace{2em}
\subsection{Width distribution of slow and fast CMEs coming from ARs and PEs}
In previous sections, we note that the slow and fast CMEs follow different power laws. This lights on the fact that may be the slow and fast CMEs have different physics involved in the mechanism that leads to their expansion and hence their widths. Since Lorentz force is responsible for propelling and expanding a CME (refer Section \ref{sec1}), we expect to see its imprint in the width distribution of slow and fast CMEs originating from different source regions.

%\begin{figure}[ht!]
%\centering
% \includegraphics[scale=0.6]{c7f2.pdf}
%  \caption{Width distribution of slow and %fast CMEs using the CACTus catalog.}
%   \label{c7f2}
% \end{figure}

\begin{table}[htb]
%\begin{center}
\footnotesize
\centering
\caption{Power-law indices of width distribution of fast and slow CMEs from different source regions obtained using two different methods}  
\label{table2}
\begin{tabular}{cc||c|cc|cccc||c|cc|cccc}
%\begin{tabular}{cccccccccccccccc}
\hline
\multicolumn{1}{c}{Source} & \multicolumn{1}{c||}{Total} & \multicolumn{7}{c||}{Fast CMEs} & \multicolumn{7}{c}{Slow CMEs} \\
%\hline
  Regions  & CMEs  &  Total & \multicolumn{2}{c|}{GF} & \multicolumn{4}{c||}{MLE} & Total & \multicolumn{2}{c|}{GF} & \multicolumn{4}{c}{MLE}\\
 \hline
      &   &     & $W_{m}$ & $\alpha$ & $W_{m}$ & $\alpha$ & $W_{d}$ & $\alpha_{d}$ &   & $W_{m}$ & $\alpha$ & $W_{m}$ & $\alpha$ & $W_{d}$ & $\alpha_{d}$ \\

ARs & 694 & 291 & 30 &-1.29 & 30 & -1.23 & $-$ & $-$ & 403 & 60 & -3.20 & 30 & -1.91 & 60 & -3.11\\
PEs & 281 & 89  & 60 &-3.43 & 30 & -2.00 & 60 & -3.74 & 192 & 60 & -3.53 & 30 & -2.08 & 60 & -3.75\\
\hline

\end{tabular}

%\end{center}
\end{table}
The entries in Table~\ref{table2} have been computed by taking events from different phases of cycle 23 and 24 as mentioned in Section 2.4. After segregating the source regions, their width distribution is studied separately. Here we use a similar power-law fitting to the width distribution of CMEs coming from the two source regions. We do power-law fitting by graphical fitting and also by MLE to estimate power-law indices.
\begin{figure}[ht]
\centering
 \includegraphics[scale=0.6]{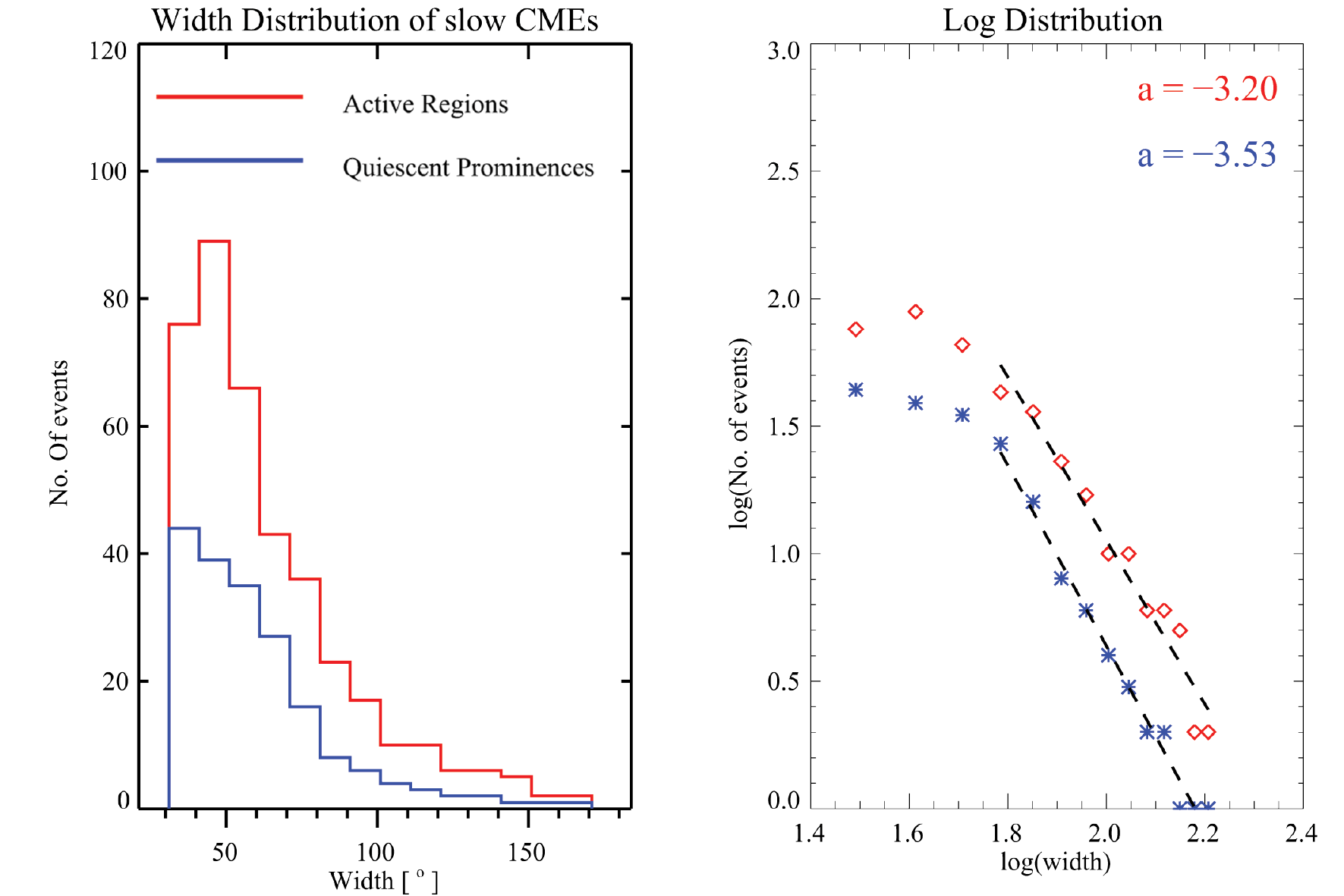}
  \caption{(Left)Width distribution of slow CMEs from different source regions, with their widths taken from the CDAW catalogue. (Right) power-law fitting of the width distribution. The black line is the power-law fit to the data. }
   \label{slow_width}
 \end{figure}
 \begin{figure}[ht]
\centering
 \includegraphics[scale=0.6]{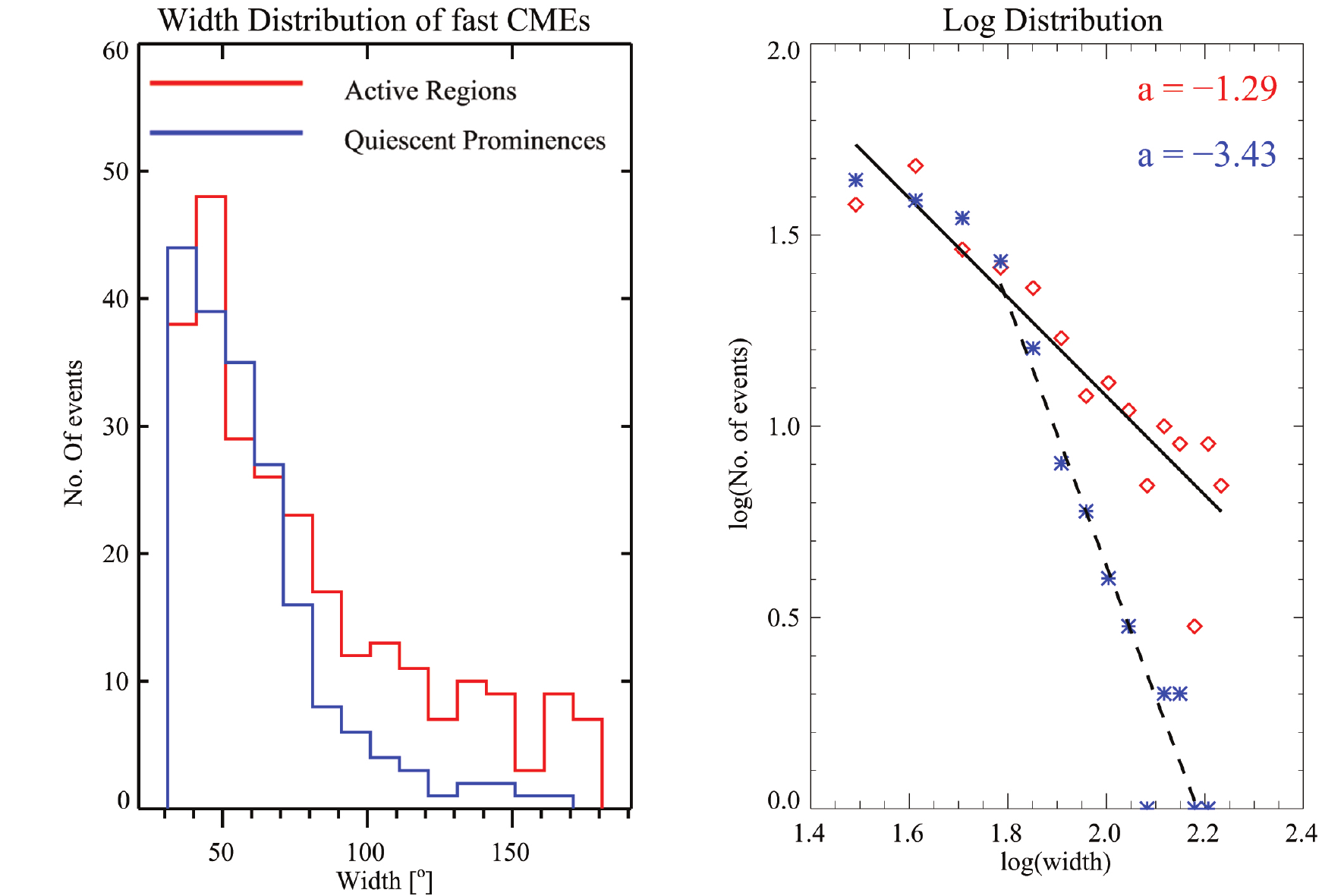}
  \caption{(Left)Width distribution of fast CMEs from different source regions, with their widths taken from the CDAW catalogue. (Right) power-law fitting of the width distribution. The black line is the power-law fit to the data.}
   \label{fast_width}
 \end{figure}
 
 Figures~\ref{slow_width} and ~\ref{fast_width} show the width distribution of slow and fast CMEs originating from ARs and PEs, and plotted alongside it are the power-law fit (in black). We get $\alpha$ as -1.29 and -3.43 for fast CMEs coming from ARs and PEs with a KS distance of 0.13 ( p-value 0.99) and 0.22 (p-value 0.86) respectively. Thus we see that indeed the width distribution of fast CMEs have different power indices for CMEs originating from ARs and PEs. In the case of slow CMEs we get $\alpha$ as -3.20 and -3.53 for CMEs from ARs and PEs with a KS distance of 0.14 (p-value 0.99) and 0.21 (p-value 0.86) respectively. Thus for slow CMEs too the power indices are slightly different for CMEs from ARS and PEs.  Again, GF is not the best method to estimate the power-law since the number of data points is small here, by using MLE fitting, we get $\alpha$ as -1.23 and -2.00 for fast CMEs coming from ARs and PEs respectively, whereas for slow CMEs we get $\alpha$ as -1.91 and -2.08 for CMEs from ARs and PEs keeping $W_{m}=30^{\circ}$ (refer, table~\ref{table2}). The 1-sigma error involved for power-law fitting of width of fast CMEs from ARs and PEs are 0.22 and 0.24 respectively, and that for slow CMEs from ARs and PEs are 0.14 and 0.03 respectively. Table~\ref{table2} clearly depicts that the power-laws followed by slow and fast  CMEs coming from ARs and PEs are distinctly different and thus supports our earlier conjecture. 
 %Further, the width distribution of CMEs coming from PEs are seen to have a steeper power-law index than those coming from ARs, irrespective of them being fast or slow. 
We note that although the slow CMEs are seen to follow a steeper power-law than the fast ones (as we have found earlier), the difference in power law indices is more pronounced for the case of CMEs associated with the ARs. Furthermore, CMEs associated with prominence eruptions and slow CMEs have steeper power index than fast CMEs coming from the active regions. The power index of fast CMEs matches with those estimated for flares. Thus it is evident that possibly the mechanism involved in the width expansion of slow and fast CMEs are different for the CMEs originating from ARs and PEs. We also note that the p-value for the power law fit to the width distribution for CMEs from different sources is lower, and this is due to the lower statistics that we have in each cases.

Thus the fact that slow and fast CMEs from ARs and PEs following different power laws in their width distribution vividly points towards a possibly different mechanism that leads to the width expansion of these CMEs and hence demands a more deeper understanding of the same. It should be noted that a few studies \citet{Zuccarello2014ApJ, Seaton2011ApJ, Kane2019ApJ} have tried to explore the possible mechanims of the CMEs but these studies are confined to isolated cases of CMEs. Also, a few statistical studies on large number of CMEs suggested that there may be different classes of CMEs with different driving mechanisms \citep{stcyr1999JGR, 2001ApJ...561..372S, Moon2002ApJ}.

From Figures~\ref{slow_width} and~\ref{fast_width} we also note that in case of the fast CMEs, effect of change in source region is more pronounced unlike the case for slow events. The limit of the field strength in the source region could be the physical reason for the strength of field in ARs, which in turn determines the available energy to power the eruptions, whereas in case of the CMEs from PEs, the magnetic structure in the periphery of the ejection site controls the chances of ejection and eventually the observed kinematic properties of the ejected material \citep{2017arXiv170903165G}. So, possibly a different mechanism governs the width expansion of CMEs coming from ARs and PEs, irrespective of them being slow or fast.

\section{Summary and Conclusions}\label{discussion}
In this work, we study the width distribution of CMEs that occurred during different phases of solar cycle 23 and 24. The CMEs were then segregated into slow ($\le$ 300 km s$^{-1}$) and fast CMEs ($\ge$ 500 km s$^{-1}$) based on the average solar wind speed, and then their width distribution was studied. We further associate the slow and fast CMEs to the source regions they originated from, and classified the identified source regions into two broad categories, ARs and PEs. We investigate if the source regions have any imprint on the width distribution of these slow and fast CMEs. The data from the CDAW catalogue has been used throughout this study. In the following, we conclude our main results from this work.
\begin{enumerate}
\item CMEs excluding `very poor' events from the CDAW catalogue tend to follow a power law in their width distribution with a power law index of -1.9 (Figure \ref{width_all_events}). Using MLE, we find the power law index to be -1.6. This power law index remains unchanged on the exclusion of the intermediate events from our sample set (Figure \ref{width_all}). Thus the intermediate events do not affect our results and thus we removed them from our sample set, as they cannot be strictly considered either slow or fast. Using GF method, we note that a single power law is unable to explain the observed distribution.
%While the power index of fast CMEs is comparable to that of the flare energy distribution \citep{1993SoPh..143..275C}; the power index of slow CMEs is too steep to be explained in terms of flare energy distributions. We also did analysis for the CMEs with width between 30--120$^{\circ}$ and found that the slow and fast CMEs have different power indices.
%We find a similar behaviour using both manual (CDAW) and automated (CACTus) catalogs. Thus, we think that it is not due to difference in the measurements made in different catalogs. It should be borne in mind that there is still a propensity for the very fast, shock-driving CMEs from the limb to appear as wide, since the shock appears as a much larger density enhancement surrounding the CME but the population of such CMEs is very small \citep[see Figure~2 in][]{2014GeoRL..41.2673G}.

\item We find different power indices for the width distribution of fast and slow CMEs (see Table~\ref{table1}). To reduce the projection effects from our results, we study the width distribution of slow and fast limb CMEs, and we found that they follow different power laws and the results remain unchanged. However, the absolute value of power indices are not the same as compared to limb and non limb CMEs which may be due to poor statistics, as demonstrated by KS test. Since both fast and slow limb and non-limb CMEs follow different power-laws in width distribution, we believe that slow and fast CMEs may have different energy sources and generation mechanisms. 

\item We study the width distribution of slow and fast CMEs coming from different source regions (ARs and PEs), and find that the power law indices are different for CMEs coming from ARs and PEs (refer Table~\ref{table2}). Furthermore, CMEs coming from PEs tend to follow a steeper power law irrespective of their speeds. Also, we find that slow CMEs tend to follow a steeper power law than fast CMEs, irrespective of the source region they are coming from. This clearly hints towards a possibly different mechanism for width expansion of these CMEs.

Thus we find from this study that apart from their speeds, slow and fast CMEs are also distinctly different in terms of the distribution of their angular width in each case. We believe that this study will help in a better understanding of the mechanism of width expansion of slow and fast CMEs coming from different source regions, and in establishing the width of a CME as a crucial parameter in the study of kinematics of CMEs and their ejection mechanisms. Extending this work on a larger sample of CMEs using de-projection methods will further help in better confirmation of our conclusions.

\end{enumerate}

\section*{Conflict of Interest Statement}

The authors declare that the research was conducted in the absence of any commercial or financial relationships that could be construed as a potential conflict of interest.

\section*{Author Contributions}
VP planned the study and performed initial analysis. SM, RP generated source region catalog. VP and SM wrote manuscript. AC performed analysis on the limb events. VP, SM, DB, and NG interpreted the results. All authors took part in the discussion.

\section*{Funding}

\section*{Acknowledgments}
We thank referees for their comments which has improved the presentation.VP is supported by the Spanish Ministerio de Ciencia, Innovaci\'on y Universidades through project PGC2018-102108-B-I00 and FEDER funds. VP was further supported by the GOA-2015-014 (KU Leuven) and the European Research Council (ERC) under the European Union's Horizon 2020 research and innovation programme (grant agreement No 724326). The authors thank IAC and IIA for providing the required computational facilities for doing this work. We also thank Mr. Bibhuti Kumar Jha for his useful comments during the work.

\section*{Data Availability Statement}
The datasets [ANALYZED] for this study can be found in the [CDAW catalogue] [\url{https://cdaw.gsfc.nasa.gov/CME_list/index.html}].
% Please see the availability of data guidelines for more information, at https://www.frontiersin.org/about/author-guidelines#AvailabilityofData

\bibliographystyle{frontiersinSCNS_ENG_HUMS} % for Science, Engineering and Humanities and Social Sciences articles, for Humanities and Social Sciences articles please include page numbers in the in-text citations
\bibliography{references}

%%% Make sure to upload the bib file along with the tex file and PDF
%%% Please see the test.bib file for some examples of references

\section*{Figure captions}

%%% Please be aware that for original research articles we only permit a combined number of 15 figures and tables, one figure with multiple subfigures will count as only one figure.
%%% Use this if adding the figures directly in the mansucript, if so, please remember to also upload the files when submitting your article
%%% There is no need for adding the file termination, as long as you indicate where the file is saved. In the examples below the files (logo1.eps and logos.eps) are in the Frontiers LaTeX folder
%%% If using *.tif files convert them to .jpg or .png
%%%  NB logo1.eps is required in the path in order to correctly compile front page header %%%

%%% If you are submitting a figure with subfigures please combine these into one image file with part labels integrated.
%%% If you don't add the figures in the LaTeX files, please upload them when submitting the article.
%%% Frontiers will add the figures at the end of the provisional pdf automatically
%%% The use of LaTeX coding to draw Diagrams/Figures/Structures should be avoided. They should be external callouts including graphics.

\end{document}